\documentstyle[12pt,aasms4]{article}
\begin{document}
\def\arcsec{\ifmmode^{\prime\prime}\;\else$^{\prime\prime}\;$\fi}
\def\arcmin{\ifmmode^{\prime}\;\else$^{\prime}\;$\fi}
\title{The Number of Publications Used as a Metric of the NOAO WIYN Queue 
Experiment}

\author{Philip Massey, Mary Guerrieri,
and Richard R. Joyce}
\affil{Kitt Peak National Observatory, National Optical Astronomy
Observatory\altaffilmark{1}
\\ P.O. Box 26732, Tucson, AZ 85726-6732}

\altaffiltext{1}{Operated by AURA
under cooperative agreement with the
National Science Foundation.}
\begin{abstract}

We use the number of papers published in 1998 and 1999 to test the
hypothesis that the queue observing mode at WIYN leads to a
significantly higher scientific throughput than classical mode
observing.  We use the papers published from the 4-m, and papers
published from the non-queue WIYN time as controls, requiring only that
the data be obtained after 1996 August 1, at which time the WIYN queue
was in its third full semester of operation, and the WIYN instruments
functional and stable.  The number of papers published from the queue
data is actually 1.5 times smaller (on a per night basis) than from the
4-m, and roughly comparable to (but lower than) the number published
from non-queue WIYN time.  Thus neither comparison offers any {\it
support} for the hypothesis that queue leads to a higher scientific
throughput.  The number of papers is relatively small, but the
statistics are sufficiently robust to {\it reject} the possibility that
queue observing at WIYN leads to a factor of 1.5 enhancement in
publication rate with a 99.3\%  confidence in comparison to the 4-m,
and with an 89.9\% confidence in comparison with non-queue WIYN time.
We consider several explanations, and urge that other observatories
planning to employ the queue mode include some controls to provide
an objective evaluation of its success. \end{abstract}


\section{Introduction}

The 3.5-m Wisconsin-Indiana-Yale-NOAO (WIYN) telescope was dedicated on
1994 October 15, and shared-risk observing began in 1995 March.  NOAO's share
of the time is 40\%, and nearly all of this has been carried out in ``queue" mode, where the observations 
from highly ranked proposals are placed in a queue and executed during
nights assigned to the queue program.  The observations are carried out by
highly experienced professionals, who are extremely familiar with
the instrumentation, without the direct assistance of the proposing astronomer.
A small fraction of the NOAO time
is scheduled out in ``classical mode", with the
observers present at the telescope.  The time allocated to the
university consortium members (roughly 60\%) 
is all carried out in classical mode.

The goal
of the NOAO WIYN queue experiment was 
eloquently described by Silva \& De Young (1996) as an empirical test
of ``the hypothesis that in the face of a high over-subscription rate, the
science throughput of WIYN can be maximized by executing
the most highly ranked science programs first, completing datasets in a timely
manner, allowing a larger range of program lengths, and matching the
observing program to the observing conditions on an observation-by-observation
basis."  

The WIYN queue has often been described as an ``experiment" at least in part
because other observatories are considering scheduling some or all of their time
in this mode, and NOAO staff have felt that what we can learn from the WIYN
queue will be useful to others.
 In an era that sees both the proliferation of very large
($\ge$8 m) telescopes, but ever-tightening financial resources, observatories
are scrambling to understand how to maximize their scientific return.

Queue observing offers a variety of theoretical advantages, as nicely
summarized by Mountain (1996) and
Boroson et al.\ (1998). For very highly ranked programs that require
rare conditions, queue observing may be the only practical way to acquire such data.  Queue observing naturally allows synoptic observations, and such
scheduling easily accommodates target-of-opportunity requests, such as optical
follow-ups of gamma-ray bursts or supernovae. Furthermore, as instrumentation
becomes more complex, queue observing carried out by dedicated observers
may result in more efficient use of telescope time than if the observations
were carried out by visitors who uses the equipment only occasionally.
This contention is partially supported by evidence that 
observers collect less data on the first night of an observing run than
on subsequent nights (Bohannan 1998).

However, there are obvious down-sides to the queue mode.  The astronomer is not
present at the telescope, and therefore cannot make real-time decisions
concerning the data.  Serendipity is eliminated, as are the risky programs 
many of us have snuck in during gaps in our main observing program, and which
have sometimes led to the more interesting results.
Some of us suspect, rightly or wrongly, 
that we could better carry out our own observations.
And, there is not the same strong sense of
``data ownership" that comes
with having carried out the observations ourselves:  the memory of a night
may provide details that are relevant to the interpretation of the reduced
data, as well as providing an emotional impetus for seeing the project
through to its completion.  

There is also a non-negligible expense of
running a queue, which is off-set to some degree
by the smaller support
required for visiting astronomers.

Boroson (1996) has described a simulation program that can be used to
test how successfully programs are completed in a queue mode
vs.\ a classical mode, using Monte Carlo sampling of characteristic 
observing conditions (weather, seeing) for the site. Boroson 
et al.\ (1998) used this simulation program comparing queue mode and classical
scheduling 
for two actual semesters (1997) of WIYN programs, concluding that 
queue scheduling at WIYN has led to a significant gain in efficiency
and scientific effectiveness.

Now that the queue experiment has run for several years, we thought it would be worth examining the gain using some real-world measure.
As emphasized by Boroson et al.\ (1998), much of the argument about observing
modes can be emotional.  We seek some metric that we can use to {\it test}
the {\it hypothesis} enunciated above that the queue observing mode leads
to significant improvement in the {\it science throughput}.  One such
simple metric is the number of refereed papers published.  This may not be
as meaningful in its long-term impact on astronomy as, say, 
the number of important new discoveries, 
but at least it has the advantage of being quantifiable, and, if the
experimental and control samples are well matched, equitable and fair.

We choose to compare the number of papers produced by the WIYN queue to the 
following two controls, 
each with its advantages and disadvantages:
\begin{enumerate}
\item The number of papers produced by observations made over the same time
period with the Mayall 4-m telescope.
\item The number of papers produced by observations made over the same time
period by non-queue use of WIYN; i.e., primarily the time used by the 
consortium universities.

\end{enumerate}

The first comparison has the primary advantage that both the 4-m and WIYN
proposals have undergone similar scrutiny by
the same time allocation committees (TACs), 
which often consider such factors as the past track-record of the proposers 
as well as the 
scientific excellence of the proposals.  Thus proposers to the 4-m and
WIYN will feel similar pressures to publish in a timely manner, and the
feasibility of the proposals has been carefully evaluated.  Users of the
university time may choose to undertake 
longer-term projects, leading ultimately
to more important results, but not processing the same rapid turn around
from observing to publication.  
We offer the second comparison as there may be differences in the actual
on-sky performance of the two telescopes that would affect the results:
the 4-m is a mature telescope, possibly with fewer teething problems, than
the newer WIYN.
 
If the queue leads to significantly higher 
scientific throughput, then we expect that the number of papers published using
data obtained via the queue should be significantly
greater than those produced by the
control samples, after normalization on the basis of the number of scheduled
nights.

\section{The Data Set}

All of the 1998 and 1999 issues of the main US astronomy journals
were examined
for papers which used 4-m and/or WIYN observations.  The complete list of 135
papers is given in Table~A1 of the Appendix.

In order to make a fair 
comparison, we restricted ourselves to only those papers for which the data
were obtained in semester ``1996B" or later (i.e., after 1996 August 1).
This was the third full semester of WIYN queue time, and the first semester in
which both the imager and fiber positioner were fully functional. 
(A non-linearity problem with the S2KB imager chip was discovered and
fixed during the 1996A semester, and a mechanical problem which compromised the
positioning accuracy of the Hydra fiber positioner was fixed in 1996 March.)

We list in Table~1 the number of papers published during 1998 and 1999.
Six papers used both 4-m {\it and} WIYN data; we chose to count each of
these papers separately for both telescopes, depending upon the date
in which the data were obtained for the telescope under consideration; i.e.,
if the data for WIYN was obtained in 1996B or later, but the 4-m data was
obtained prior to 1996, it would count as a WIYN publication but not as a
4-m paper. There were six papers in our list in which the data
collected were such a minor component of the
paper that we chose not to count the paper at all; only one of these used
the WIYN queue, and in that case the data had been published previously by
the original proposers.

\section{Results}

\subsection{Comparison of the WIYN and the 4-m}

In order to make a valid comparison, we must first take into account that not
as much time is scheduled for the WIYN queue as for the 4-m.  We expect the
answer is about 40\%, as NOAO receives 
40\% of the time on WIYN, and almost
all of this goes to the queue.  However, the 4-m is shut down during July
and August, while WIYN continues to operate; on the other hand, there
are more engineering nights scheduled at WIYN. One could use the 
total number of clear hours spent observing as
the normalization, but these data are hard to extract reliably.  Instead, we
took the final observing schedules for semesters 1996B, 1997A, 1997B, 1998A, and
1998B, and simply counted the number of nights assigned to the WIYN queue,
and to science operations at the 4-m. (For the latter, we included half-night
instrument ``checkout" nights, as much of this time is typically 
returned to the observers
scheduled on the second half; full-night ``check" nights and engineering
nights were excluded.  We excluded all engineering nights scheduled at WIYN,
although occasionally queue observations are obtained during such time.)  
The numbers of nights so scheduled for the WIYN
queue and for the 4-m are 260 and 656 respectively; i.e.,
the number of nights scheduled to the WIYN queue turned out to be 39.6\% of the
nights scheduled at the 4-m.

If the hypothesis described above is
correct, we would expect the number of publications based upon WIYN queue data
to be significantly greater than 40\% of those produced by the 4-m.
Instead, we find in Table~1 that there
were only 9 papers produced by WIYN queue data as opposed to 34 papers
produced by the 4-m; i.e., 26\%.  Thus there are actually
1.5 times fewer papers published (on a per night basis) based on
queue WIYN data relative to those based on 4-m data. This comparison does not
support the hypothesis of greater science throughput by the WIYN queue.

Can we rule out the hypothesis given the small number statistics?  If we assume
the simplest model that a 1$\sigma$ uncertainty in the number of publications
$N$ is simply the $\sqrt{N}$, then the 1$\sigma$ error on the 0.26 ratio of
WIYN to 4-m publications is 0.13.  What does it mean for there to be a 
``significant" enhancement in the scientific throughput?  Boroson et al.\ (1998)
discuss how their simulation predicts this will depend upon program type,
TAC grade, and so on, and that overall about 2.5 times as many programs will be completed by queue observing than with classical observing.  We take here
a more conservative approach: certainly a 50\% increase (a factor of 1.5)
would be cause for celebration. Were this enhancement present, we would expect
there to be 1.5 $\times$ 39.6\% = 59.4\% as many WIYN queue papers as 4-m
papers.  We observe 0.26$\pm$0.13
We thus can reject such an increase at a +2.5$\sigma$ level; i.e., with a 99.3\%
confidence.\footnote{The rejection probability corresponding to +2.5$\sigma$
was found by  $$1.0-0.5 \times (1.0-A_G(\mid x-\mu\mid /\sigma)),$$ where  
$A_G$ is the
integral probability of the normal distribution with a mean of $\mu$ and
a standard deviation of $\sigma$; see, for example, Fig.~C-2
in Bevington (1969).}

\subsection{Comparison of Queue vs.\ Non-Queue Time at WIYN}

Of the 731 nights scheduled for science at WIYN during 1996B through 1998B,
we find that 260 nights were scheduled for queue observations (35.6\%),
27 nights were scheduled for NOAO classical observations (3.7\%), and 444 as
university time (60.7\%).  If queue observing produced
a significantly higher scientific throughput, we would expect significantly
more than 36\% of the papers produced by WIYN data to be based on data obtained
with the queue.  Instead, of the 28 total WIYN
papers in our sample, 9
(32\%) were produced from queue data. This is 
essentially
the same fraction of time on WIYN used by the queue (36\%), and therefore does
not suggest that queue provides a significant advantage. 

While the data fail to offer any support for the hypothesis, at what level
can we reject the claim, given our limited statistics? Using the same argument
as above that we would hope for a factor of 1.5 enhancement over the non-queue
publication rate, we can ask at what level can we exclude the queue publications
amounting to 1.5$\times$ 35.6\% = 53.4\% of the total.  The uncertainty
in our ratio 0.32 ratio is 0.17.  Thus we can exclude a 50\% enhancement at
the +1.3$\sigma$ level; i.e., with an 89.8\% confidence. 

Nevertheless, it is clear that queue observing does
fare better in this comparison than it did in comparison to the 4-m control,
although still failing to produce a higher number of publications. 
Several explanations
come to mind.
One possibility is that the 4-m simply operates more efficiently
than WIYN (at least in the time period when most of the data were acquired), 
and that
it was thus easier to obtain usable data at the 4-m.
It is possible that review of queue proposals by an
outside TAC leads to a higher publication rate than time used by the
universities, who have a preallocated amount of time, which is divided up
internally.  (As suggested earlier, the university time may be spent on 
longer-term programs than the NOAO portion.)  Finally, the 4-m supports a wider
complement of instrumentation (such as infrared imaging and spectroscopy) than WIYN,
which plausibly provides greater coverage of astronomical disciplines and
thus involvement in a wider variety of publications.

Although the numbers are small, the very high publication rate
for NOAO time that is scheduled {\it classically} at WIYN suggests that it
may be the TAC process rather than the telescope or instrumentation
which explains why the queue
does better in this comparison than it does in comparison to the 4-m: 
14\% of the WIYN papers were produced by the small
(3.7\%) time allocated to non-university classical observing. 
The classically scheduled NOAO time 
undergoes the same rigorous review as the queue
proposals, and thus is under the same pressure to publish rapidly.

\section{Discussion}

Arguably, the WIYN queue has been as well run as it is possible for
any queue to be.  A survey carried out of astronomers who had proposed for
queue time suggests that people were very satisfied with the quality of the
data they received (Boroson et al.\ 1998); some might expect that maintaining
data quality to be the hardest part of a queue.
Yet the evidence so far fails to support the suggestion that queue
observing leads to a higher scientific throughput, at least as measured by the
number of publications.  Why does this differ from the dramatic predictions of simulations that suggest that a much higher percentage of programs should be
completed by the queue mode? 

We have read through the papers based upon the WIYN queue data and have several observations of our own to offer.  First, let us consider the advantage that
queue offers in providing easy ``target of opportunity" (TOO) observations.
Of the full set of 11 papers (ignoring the 1996B cutoff), four rely on the
TOO advantage of queue for optical followup of 
gamma-ray bursts (Galama et al.\ 1998) or supernovae
(Jha et al.\ 1999; Perlmutter et al.\ 1999; and Riess et al.\ 1998).  
Although WIYN played a role in these important studies, our examination of
these papers suggests that it was a relatively minor role, with the
majority of the data coming from elsewhere.   For instance,
there are considerably
more
data from the CTIO 4-m (which is classically scheduled) than from WIYN
in the Riess et al.\ (1998) study.
Inspection of these papers suggest that
there is no lack of ways for large groups to acquire such data. 
The number of authors on these four papers range from
17 to 42, and with a large number of participants being a reflection of
the degree (and method?) of telescope access. 
Thus TOO use of WIYN may not be more significant simply
because there
are other ways of obtaining such data.

One of the other purported advantages for queue observing is the ability to 
take
advantage of particularly good
conditions, and indeed some programs may not be
completed any other way.  However, this advantage is larger the greater the
range of conditions. 
For instance, if the frequency histogram of delivered image quality (DIQ)
is very sharply peaked,
then queue offers less of an advantage, as all programs will obtain something
like the median seeing.  At WIYN the median DIQ (at {\it R})
is
0.8 arcseconds, and 0.6 arcsecond or better images are achieved 18\% of the
time (Green 1999).  
Of the 11 queue papers listed in Table~A1, 
Armandroff, Jacoby, \& Davies (1999) is one of the clearest examples of taking
advantage of the queue to obtain the best DIQ. 
The study utilized sub-arcsecond conditions (0.8 arcsec at {\it B},
0.6 arcsec at {\it V}, and 0.7 arcsec at {\it I}) for deep imaging of a newly
discovered dwarf member of the Local Group, Andromeda~VI, after confirming its
nature using imaging at the 4-m.  Nevertheless, these DIQ 
values are not all that
different than the median values.

However, it may be that the sociological issues raised in the introduction dominate.  The
use of queue may reduce the sense of ``data ownership," and given
situations of ``data saturation," we are more likely to publish the
data more rapidly if we have acquired them ourselves.  
The use of ``queue mode" on {\it HST} has been perceived as being highly
successful, 
although a meaningful control sample is hard to find for comparison; however,
one important difference comes to mind, namely that observing time (to US
proposers) usually comes with grants, providing a financial incentive to produce
results rapidly, coupled with a 1-year proprietary
period for unique data.  An additional consideration is that {\it HST} supplies
the user with fully reduced data, unlike WIYN, which provides basic calibration
data and requested standard observations, but which does not attempt a 
``pipe-line" reduction. However, our own experience with {\it HST} data is
that customized reductions are often needed in order to provide the data most
meaningful for a particular application.

Finally, it may be that we simply have not been sufficiently patient.  As is evident from the 4-m publications, only one-third of the 4-m papers in
the past two years relied purely on ``new" data (i.e., all data obtained in
the past 3.5 years).  While our control samples explicitly took this into account,
we are nevertheless comparing numbers that are on the
the tails of the distribution of how quickly data finds its way into the
literature.  This may be particularly true if the datasets from the WIYN
queue were to be larger than that in the control samples, or if they
take longer to reduce.
Current plans call for discontinuing the WIYN queue at the
end of semester 2000A, but continuing to provide some synoptic and target of opportunity service observing beyond that.  It will be interesting to
re-examine the literature five years from now
using data obtained in 1996B-2000A
as the selection criterion.

We note that the quantity we would most like to measure is ``quality", but
this is of course harder to do in an objective manner.  Citation rates might
provide one means, but not enough time has past for these to be meaningful.
Counting the number of papers is some measure of the ``output" of a telescope,
but it is not necessarily the best; it does have the advantage of being
objective and reproducible, qualities usually assumed to be desirable
in any experiment.

Nevertheless, our results suggest that it 
may benefit observatories to evaluate
their queue programs using some external measure, such as the number of
publications, if suitable controls can be defined.

\acknowledgments
Helmut Abt, 
Dave De Young, David Sawyer, 
Dave Silva, and Sidney C. Wolff were kind enough to 
provide thoughtful comments
on the manuscript.  We also benefited conversations with Taft Armandroff,
Bruce Bohannan, and
Abi Saha on the issues of queue observing.

\section{Appendix}

In Table~A1 we present the list of papers published in the
{\it Astronomical Journal}, the {\it Astrophysical Journal} (Parts 1 and 2), 
and the {\it Publications of the Astronomical Society of the Pacific}
during 1998 and 1999 that used data from the 4-m and/or WIYN.  We list
the dates of the first data obtained (from the relevant telescope).  Often
this information was directly obtained from the paper, but in many cases we
had to contact the authors, or inspect the observing schedule or list of
queue programs to determine the actual data or semester.

\end{document}


\setcounter{page}{0}
\pagestyle{empty}
\begin{deluxetable}{l l l l l}
\tablewidth{475pt}
\tablenum{A1}
\tablecaption{Papers Published in 1998+1999 Using 4-m and/or WIYN Data}
\tablecolumns{5}
\tablehead{
\colhead{Citation}
& \multicolumn{4}{c}{Date of First Data\tablenotemark{a}} \\ \cline{2-5}
\colhead{}
&\colhead{4-m Mayall}
&\colhead{WIYN--Queue}
&\colhead{WIYN--NOAO Classical}
&\colhead{WIYN--University}
}
\startdata
\sidehead{\it Astronomical Journal}
 115, 436    & \nodata   & \nodata & \nodata & 1995 Jul \nl
 115, 535   & ?\tablenotemark{b} & \nodata & \nodata & \nodata \nl
 115, 573   & \nodata    & \nodata & \nodata & 1995 Oct \nl
 115, 975    & \nodata   & \nodata & \nodata & 1996 Oct \nl
 115, 1329   & 1996 Apr    & \nodata   & \nodata & \nodata \nl
 115, 1869  &  \nodata   & \nodata &  \nodata  & 1996 Sep \nl
 115, 2018  & \nodata     & \nodata & 1997 July & \nodata \nl
 115, 2044 & \nodata     & \nodata  & \nodata  & 1996 Mar \nl
 115, 2059   & 1991      & \nodata   & \nodata & \nodata \nl
 115, 2074 &  1998 Jan   & \nodata   & \nodata & \nodata \nl
 116, 102       & 1994 Mar    & \nodata   & \nodata & \nodata \nl
 116, 146    & 1996 Oct  & \nodata   & \nodata & \nodata \nl
 116, 455    & 1996 Jan  & \nodata   & \nodata & \nodata \nl
 116, 549    & 1996 Oct   & \nodata   & \nodata & \nodata \nl
 116, 673   & 1994 Nov    & \nodata & \nodata & \nodata \nl
 116, 707    & 1993 Nov     & \nodata   & \nodata & \nodata \nl
 116, 1066   & 1996 Apr  & \nodata   & \nodata & \nodata \nl
 116, 1074   & 1992 Oct  & \nodata   & \nodata & \nodata \nl
 116, 1221  & 1998 May    & \nodata & \nodata & \nodata \nl
 116, 1367  &  \nodata   & \nodata & 1997 Apr & \nodata \nl
 116, 1396   & 1997 Oct    & \nodata   & \nodata & \nodata \nl
 116, 1412   & 1996 Jan  & \nodata   & \nodata & \nodata \nl
 116, 1789       & \nodata   & \nodata & \nodata & 1995 Jul \nl
 116, 2287  &  1997 Oct   & \nodata & \nodata   & \nodata \nl
 116, 2984  & 1998 Jun   & \nodata & \nodata & \nodata \nl
 117, 75    & \nodata     & \nodata & \nodata & 1997 May \nl
 117, 181    & \nodata   & \nodata   & \nodata & 1997 Apr \nl
 117, 308   & $<$1996B   & \nodata & \nodata & \nodata \nl
 117, 548   & \nodata    & \nodata & \nodata & 1997 \nl
 117, 1023   & 1993      & \nodata   & \nodata & \nodata \nl
 117, 1402  &  1996 Apr   & \nodata   & \nodata & \nodata  \nl
 117, 1890   & 1982      &\nodata    & \nodata & \nodata \nl
 117, 2077  & \nodata    & \nodata & \nodata & 1997 \nl
 117, 2666   & 1994 Mar   & \nodata   & \nodata & \nodata \nl
 117, 2919  & 1997 Oct    & \nodata & \nodata & \nodata \nl
 118, 76     & 1993    & \nodata   & \nodata & \nodata \nl
 118, 777   & 1998 May    & \nodata & \nodata & \nodata \nl
 118, 236    & 1989   & \nodata   & \nodata & \nodata \nl
 118, 509 & 1982 & \nodata & \nodata & \nodata \nl
 118, 633    & 1998 Jun & \nodata   & \nodata & \nodata \nl
 118, 719   & 1993       & \nodata & \nodata & \nodata \nl
 118, 1220  &  1998 Jan   & 1998 Sept & \nodata & \nodata \nl
 118, 1542  & 1998 Jun & \nodata & \nodata & \nodata \nl
 118, 1657  & \nodata     & \nodata & \nodata & 1997 Oct \nl
 118, 1806  & 1992 Sep   & \nodata & \nodata & \nodata \nl
 118, 2002 & 1997 Apr    & \nodata & \nodata & \nodata \nl
 118, 2184  & 1998 Jan   & \nodata & \nodata & \nodata \nl
 118, 2409  & \nodata     & \nodata & \nodata &  1997 Nov \nl
 118, 2466   & 1998 Dec  &\nodata    & \nodata & \nodata \nl
 118, 2511       & $<$1982    & \nodata   & \nodata & \nodata \nl
 118, 2547   & 1997 Jul & \nodata & \nodata & \nodata \nl
 118, 2751   & \nodata      & \nodata & \nodata & 1997 Apr \nl
 118, 2775   & 1991 Nov & \nodata & \nodata & \nodata \nl
 118, 2940   & 1998 Dec & \nodata & \nodata & \nodata \nl
\sidehead{\it Astrophysical Journal}
 492, 62   & 1995 Apr  & \nodata & \nodata & \nodata \nl
 492, 461     & 1993  & \nodata   & \nodata & \nodata \nl
 492, L21   & 1996A     & \nodata   & \nodata & \nodata \nl
 493, L27  & \nodata & \nodata &  \nodata & 1997 Aug \nl
 494, L185  & \nodata   & \nodata   & \nodata & 1995 Dec \nl
 495, 100   & 1996B(?)  & 1996B     & \nodata & \nodata \nl
 495, 933  & \nodata    & \nodata & \nodata & 1996 Oct \nl
 496, 103  &  $<$1995   & 1996A     & \nodata & \nodata \nl
 496, 803  & \nodata     & \nodata & 1995 Oct & \nodata \nl
 497, 227  & 1994 Mar   & \nodata & \nodata & \nodata \nl
 497, 294   & 1996 May  & \nodata   & \nodata  & \nodata \nl
 497, L13  &  \nodata   & 1997 May & \nodata & \nodata \nl
 497, L75   & \nodata   & \nodata   & \nodata & 1997 Feb \nl
 499, 577   & 1989      & \nodata   & \nodata  & \nodata \nl
 499, 828  &  1988       & \nodata   & \nodata & \nodata \nl
 500, 173  &  1996 Apr   & \nodata & \nodata   & \nodata \nl
 500, 188  & 1994 Mar   & \nodata & \nodata & \nodata \nl
 500, L105 & \nodata    & 1997 May\tablenotemark{b} & \nodata & 1997 May\tablenotemark{b}\nl
 500, L11  & \nodata   & 1997 Oct & \nodata & \nodata \nl
 501, 137  & 1996 Dec   & \nodata   & \nodata & \nodata \nl
 501, 153  & 1996 Aug  &\nodata    & 1996 Oct & \nodata \nl
 501, 624  & 1994 Nov    & \nodata & \nodata & \nodata \nl
 502,  16  & 1995 Jun   & \nodata & \nodata & \nodata \nl
 502, L39   & 1995 Jun   & \nodata   & \nodata & \nodata \nl
 503, 109  & 1997 Mar    & \nodata & \nodata & \nodata \nl
 503, 543  & 1996 May    & \nodata & \nodata & \nodata \nl
 504, 935   & \nodata      & 1997A     & \nodata & \nodata \nl
 505, 174  & 1996 Feb    & \nodata & \nodata & \nodata \nl
 505, 199     & 1985  & \nodata   & \nodata & \nodata \nl
 505, 793   & 1996 Aug  & \nodata   & 1996 Oct & \nodata \nl
 506,  33   & 1993  & \nodata   & \nodata & \nodata \nl
 506, 222   & 1994 Apr  & \nodata   & \nodata & \nodata \nl
 507, 210   & 1991 Nov  & \nodata   & \nodata & \nodata \nl
 507,  46      & 1995 Apr \tablenotemark{b}  & \nodata   & \nodata & \nodata \nl
 507, 558  & 1994        &  \nodata & \nodata & \nodata \nl
 508, 200      & \nodata   & \nodata & \nodata & 1995 Jul \nl
 508, 397   & 1992 Oct  & \nodata   & \nodata & \nodata \nl
 509,  93   & 1996 Feb    & \nodata   & \nodata & \nodata \nl
 510,  82   & 1994       & \nodata & \nodata & \nodata \nl
 510, 197   & 1997 Oct   & \nodata & \nodata & \nodata \nl
 510, 197   & 1996 Jan    & \nodata & \nodata & \nodata \nl
 510, 251   & 1997 Sep   & \nodata & \nodata & \nodata \nl
 511, 639  & 1994 Mar   & \nodata & \nodata & \nodata \nl
 511, L1   & 1997 Apr  & \nodata & \nodata & \nodata \nl
 512, 125     & 1978\tablenotemark{b} & \nodata & \nodata & \nodata \nl
 513, 34   & 1996 Apr    & \nodata & \nodata & \nodata \nl
 513, L25   & 1997 Feb      & \nodata   & \nodata & \nodata \nl
 514, 746   & 1994    & \nodata   & \nodata & \nodata \nl
 515, 169  & 1995 Sep   & \nodata & \nodata & \nodata \nl
 515, 191   & 1995 Sep   & \nodata & \nodata & \nodata \nl
 515, 487     & 1993 Oct  & \nodata   & \nodata & \nodata \nl
 516, 563  &  1995 Mar & \nodata   & \nodata & \nodata \nl
 517,  40   &  1996A      & \nodata   & \nodata & \nodata \nl
 517, 130 & 1987 May    & \nodata & \nodata & \nodata \nl
 517, 565   & \nodata   & 1995A   & \nodata & \nodata \nl
 517, 692   & 1998 Apr    & \nodata   & \nodata & \nodata \nl
 518, 167   & 1996 Nov  & \nodata   & \nodata & \nodata \nl
 519, 610  & 1997 Jun    & \nodata & \nodata & \nodata \nl
 520, 751  & 1992 Apr    & \nodata & \nodata & \nodata \nl
 520, L103  & \nodata   & 1998 Aug    & \nodata & \nodata \nl
 521, L37  & \nodata     & \nodata & \nodata & 1995 Dec \nl
 522, 199 & \nodata     & \nodata & \nodata & 1996 Nov \nl
 522, 338   & 1997 Sep  & \nodata   & \nodata & \nodata \nl
 524, 1000 & 1998  & \nodata & \nodata & \nodata \nl
 524, L103    & \nodata   & \nodata & \nodata & 1999 Jun \nl
 525, 466   & 1998 Dec  & \nodata   & \nodata & \nodata \nl
 525, 659     &  $<$1984 & \nodata   & \nodata & \nodata \nl
 526, 274  & \nodata     & 1997 Nov & \nodata & \nodata \nl
 526, 665   & \nodata & \nodata & \nodata & 1995 Dec \nl
 527, 110   & 1994 Mar\tablenotemark{b} & \nodata & \nodata & \nodata \nl
 527, 219   & 1997 Jan  & \nodata & \nodata & \nodata \nl
\sidehead{\it Astrophysical Journal Supplements}
 114, 133 & 1992 Sep    & \nodata & \nodata & \nodata \nl
 117,  89  & 1990 Sep    & \nodata & \nodata & \nodata \nl
 118, 353  & \nodata   & \nodata & \nodata & 1996 Aug \nl
 119,   1 & 1994          & \nodata & \nodata & \nodata \nl
 119, 189  & 1991 Apr   & \nodata   & \nodata & \nodata \nl
 121, 417  & 1994     & \nodata   & \nodata & \nodata \nl
 125,  73  & \nodata    & 1998 May  & \nodata & \nodata \nl
 125, 489  & 1996 Oct   & \nodata & \nodata & \nodata \nl
\sidehead{\it Publications of the Astronomical Society of the Pacific}
 111,  685  & 1990 Sep  & \nodata & \nodata & \nodata \nl
 111, 1233   & 1993 Jun & 1996 Aug  & \nodata  & \nodata \nl
\enddata
\tablenotetext{a}{In the cases when only the observing semester could be
ascertained, we list that information.  ``A" refers to the period of 1 February
through 31 July, and ``B" from 1 August through 31 January of the following year
.}
\tablenotetext{b}{Minor use; not counted.}
\end{deluxetable}